\documentstyle[sprocl]{article}
\input{psfig}
\def\Journal#1#2#3#4{{#1} {\bf #2}, #3 (#4)}

\def\NIM{\em Nucl. Instrum. Methods}

\def\NPB{{\em Nucl. Phys.} B}
\def\PLB{{\em Phys. Lett.}  B}
\def\PRL{\em Phys. Rev. Lett.}
\def\PRD{{\em Phys. Rev.} D}

\def\be{\begin{equation}}
\def\ee{\end{equation}}
\def\bea{\begin{eqnarray}}
\def\eea{\end{eqnarray}}

\def\etal{{\it et al}}
\begin{document}
\hspace*{\fill} HEPSY 97-01\linebreak
\hspace*{\fill} June 1997~\linebreak
\centerline{~}
\centerline{~}
\title{THE BTeV PROGRAM}
\author{SHELDON STONE}
\address{Physics Department, 201 Physics Building, Syracuse Univerisity,\\
Syracuse, NY 13244-1130, USA}
\maketitle\abstracts{A brief description is given of BTeV a proposed program
at the Fermilab collider sited at the C0 intersection region. The main goals
are measurement of mixing, CP violation and rare decays in both the $b$ and
charm systems. The detector is a two arm forward-backward spectrometer capable
of triggering on detached vertices and dileptons, and possessing excellent 
particle identification, electron, photon and muon detection.}

\vspace{2.4cm}
\begin{flushleft}
.\dotfill .
\end{flushleft}
\begin{center}
{\it Presented at ``B Physics and CP Violation," Honolulu, Hawaii, March
1997; to appear in the proceedings.} 
\end{center}
\newpage

\section{Introduction}
BTeV is a Fermilab collider program whose main goals are to measure
mixing, CP violation and rare decays in the $b$ and $c$ systems. Using the new
Main injector, now under construction, the collider will produce on the order
of $10^{11}$ $b$ hadrons in $10^7$ sec. of running. This compares favorably
with $e^+e^-$ colliders operating at the $\Upsilon$(4S) resonance. These
machines, at their design luminosities of $3\times 10^{33}$cm$^{-2}$s$^{-1}$
will produce $6\times 10^7$ $B$ mesons in $10^7$ seconds.\cite{bfacs} 

\section{Importance of Heavy Quark Decays}
The physical point-like states of nature that have both strong and electroweak
interactions, the quarks, are mixtures of base states described by the
Cabibbo-Kobayashi-Maskawa matrix:\cite{ckm}
\begin{equation}
\left(\begin{array}{c}d'\\s'\\b'\\\end{array} \right) =
\left(\begin{array}{ccc} 
V_{ud} &  V_{us} & V_{ub} \\
V_{cd} &  V_{cs} & V_{cb} \\
V_{td} &  V_{ts} & V_{tb}  \end{array}\right)
\left(\begin{array}{c}d\\s\\b\\\end{array}\right)
\end{equation}
The unprimed states are the mass eigenstates, while the primed states denote
the weak eigenstates. A similar matrix describing neutrino mixing is possible
if the the neutrinos are not massless.

There are nine complex CKM elements. These 18 
numbers can be reduced to four independent quantities by applying unitarity 
constraints and the fact that the phases of the quark wave functions are arbitrary. 
These four remaining numbers are  fundamental constants of nature that 
need to be determined from experiment, like any other
fundamental constant such as $\alpha$ or $G$. In the Wolfenstein 
approximation the matrix is written as\cite{wolf}
\begin{equation}
V_{CKM} = \left(\begin{array}{ccc} 
1-\lambda^2/2 &  \lambda & A\lambda^3(\rho-i\eta) \\
-\lambda &  1-\lambda^2/2 & A\lambda^2 \\
A\lambda^3(1-\rho-i\eta) &  -A\lambda^2& 1  
\end{array}\right).
\end{equation}
The constants $\lambda$ and $A$ have been measured.\cite{virgin}

The phase $\eta$ allows for CP violation. 
CP violation thus far has only been seen in the neutral kaon 
system. If we can find CP violation in the $B$ system we could see
if the CKM  model works or perhaps go beyond the model. Speculation has it that
CP violation  is  responsible for the baryon-antibaryon asymmetry in our
section of the Universe. If  so,  to understand the mechanism of CP violation
is critical in our conjectures of why  we  exist.\cite{langacker}

\begin{figure}[hbtp]
\vspace{-6mm}
\centerline{\psfig{figure=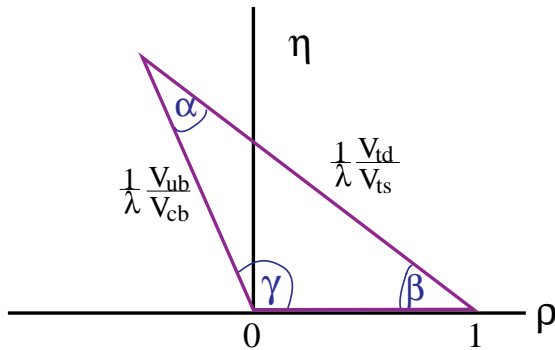,height=2.in}}
\vspace{-4mm}
\caption{The unitarity triangle shown in the $\rho-\eta$ plane. The
left side is determined by measurements of $b\to u/b\to c$ and the right side can be
determined using mixing measurements in the $B_s$  and $B_d$ systems. The 
angles can be found by making measurements of CP violating asymmetries in
hadronic $B$ decays.
\label{ut_tri}}
\end{figure}

Unitarity of the CKM matrix leads to the constraint triangle
shown in Fig.~\ref{ut_tri}. The leftside can be measured using charmless
semileptonic $b$ decays, while the rightside can be measured by using the ratio
of $B_s$ to $B_d$ mixing. The angles can be found by measuring CP violating
asymmetries in hadronic $B$ decays. 

\begin{figure}[htbp]
\vspace{-1.4cm}
\centerline{\psfig{figure=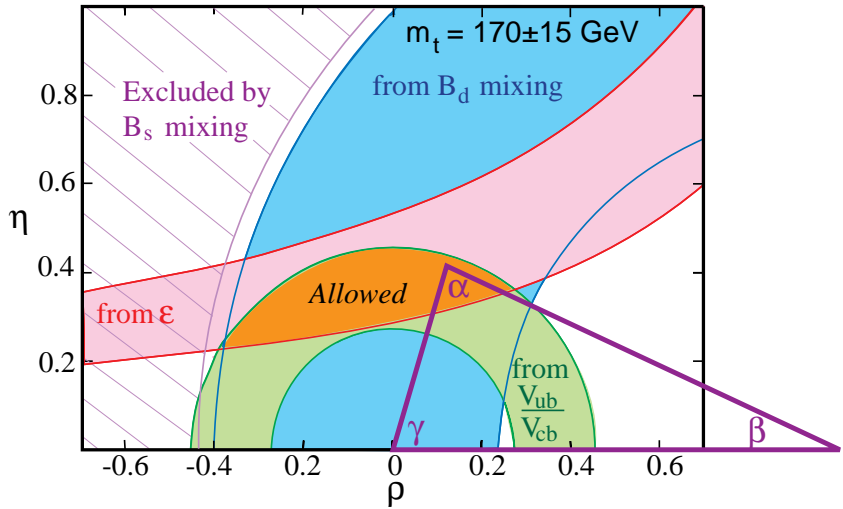,height=4.3in,bbllx=0bp,bblly=200bp,bburx=600bp,bbury=700bp,clip=}}
\vspace{-3.6cm}
\caption{\label{ckm_fig}The regions in $\rho-\eta$ space (shaded) consistent
with measurements of CP violation in $K_L^o$ decay ($\epsilon$), $V_{ub}/V_{cb}$
in semileptonic $B$ decay, $B_d^o$ mixing, and the excluded region from
limits on $B_s^o$ mixing. The allowed region is defined by the overlap of
the 3 permitted areas, and is where the apex of the  CKM triangle  sits. The
bands represent $\pm 1\sigma$ errors. The error on the $B_d$ mixing band is
dominated by the parameter $f_B$. Here the range  is taken as 
$240> f_B > 160$ MeV.}
\end{figure}

The current status of constraints on $\rho$ and $\eta$ is shown in 
Fig.~\ref{ckm_fig}. One constraint on $\rho$ and $\eta$ given by the $K_L^o$ CP
violation  measurement ($\epsilon$),\cite{buras} where the largest error arises
from theoretical uncertainty. Other constraints come from current measurements
on $V_{ub}/V_{cb}$, and $B_d$  mixing.\cite{virgin} The width of both of these
bands are also dominated by theoretical errors. Note that the errors used are
$\pm 1\sigma$. This shows that the data are consistent with the standard model
but do not pin down $\rho$ and $\eta$.

It is crucial to check if measurements of the sides and angles are consistent,
i.e., whether or not they actually form a triangle. The standard model is
incomplete. It has many parameters including the four CKM numbers, six quark
masses, gauge boson masses and coupling constants. Perhaps measurements of the
angles and sides of the unitarity triangle will bring us beyond the standard
model and show us how these paramenters are related, or what is missing. 

\section{The Main Physics Goals of BTeV}
\subsection{Physics Goals For B's}
Here I briefly list the main physics goals for studies of the $b$ quark.

$\bullet$ Precision measurements of $B_s$ mixing, both the time evolution $x_s$
and the lifetime difference, $\Delta\Gamma$, between the positive CP and
negative CP final states.

$\bullet$ Measurement of the ``CP violating" angles $\alpha$ and $\gamma$. We
will use $B^o\to \pi^+\pi^-$ for $\alpha$ and measure $\gamma$ using several
different methods including $B^+\to D^oK^+$, $B^+\to \overline{D}^oK^+$, where
the $D^o$ can decay directly or via a doubly Cabibbo suppressed decay mode. We
also need measure the conjugate $B^-$ decay modes.\cite{gronau,sad}

$\bullet$ Search for rare final states such as $K\mu^+\mu^-$ and $\pi\mu^+\mu^-$
which could result from new high mass particles coupling to $b$ quarks.

$\bullet$ We assume that the CP violating angle $\beta$ will have already been
measured by using $B^o\to \psi K_s$, but we will be able to significantly
reduce the error. 

\subsection{The Main Physics Goals for charm}
According to the standard model, charm mixing and CP violating effects should
be ``small." Thus charm provides an excellent place for non-standard model
effects to a appear. Specific goals are listed below.

$\bullet$ Search for mixing in $D^o$ decay, by looking for both the rate of 
wrong sign decay, $r_D$ and the width difference between positive CP and
negative CP eigenstate decays, $\Delta\Gamma$. The current upper limit on $r_D$
is $3.7\times 10^{-3}$, while the standard model expectation is
$r_D<10^{-7}$.\cite{dmix}

$\bullet$ Search for CP violation in $D^o$. Here we have the advantage over $b$
decays that there is a large $D^{*+}$ signal which tags the inital flavor of
the $D^o$ through the decay $D^{*+}\to \pi^+ D^o$. Similarly $D^{*-}$ decays
tag the flavor of inital $\overline{D}^o.$ The current experimental upper 
limits on CP violating asymmetries are on the order of 10\%, while the standard
model predictiion is about 0.1\%.\cite{dcp}

$\bullet$ Search for direct CP violation in charm using $D^+$ and $D_s^+$ decays.

$\bullet$ Search for rare decays of charm, which if found would signal new
physics.

\subsection{Other $b$ and charm Physics Goals}
There are many other physics topics that can be addressed by BTeV. A short list
is given here.

$\bullet$ Measurement of the $b\overline{b}$ production cross-section and
correlations between the $b$ and the $\overline{b}$ in the forward direction.

$\bullet$ Measurement of the $B_c$ production cross-section and decays.

$\bullet$ The spectroscopy of $b$ baryons.

$\bullet$ Precision measurement of $V_{cb}$ using the usual mesonic decay
modes the baryonic decay mode
$\Lambda_b\to \Lambda_c\ell^-\bar{\nu}$ to check the form-factor shape
predictions.

$\bullet$ Precision measurement of $V_{ub}/V_{cb}$ using the baryonic decay
modes $\Lambda_b\to p\ell^-\bar{\nu}$ and 
$\Lambda_b\to \Lambda_c\ell^-\bar{\nu}$  and the usual mesonic decay modes.

$\bullet$ Measurement of the $c\overline{c}$ production cross-section and
correlations between the $c$ and the $\overline{c}$ in the forward direction.

$\bullet$ Precision measurement of $V_{cd}$ and the form-factors in the
decays $D\to\pi\ell^+\nu$ and $D\to\rho\ell^+\nu$.

$\bullet$ Precision measurement of $V_{cs}$ and the form-factors in the
decay $D\to K^*\ell^+\nu$.

\section{Characteristics of Hadronic $b$ Production}

It is often customary to characterize heavy quark production in hadron
collisions with the two variables $p_t$ and $\eta$. The later variable was
first invented by those who studied high energy cosmic rays and is assigned the
value 
\begin{equation}
\eta = -ln\left(\tan\left({\theta/2}\right)\right),
\end{equation}
where $\theta$ is the angle of the particle with respect to the beam direction.

\begin{figure}[htb]
\vspace{-.4cm}
\psfig{figure=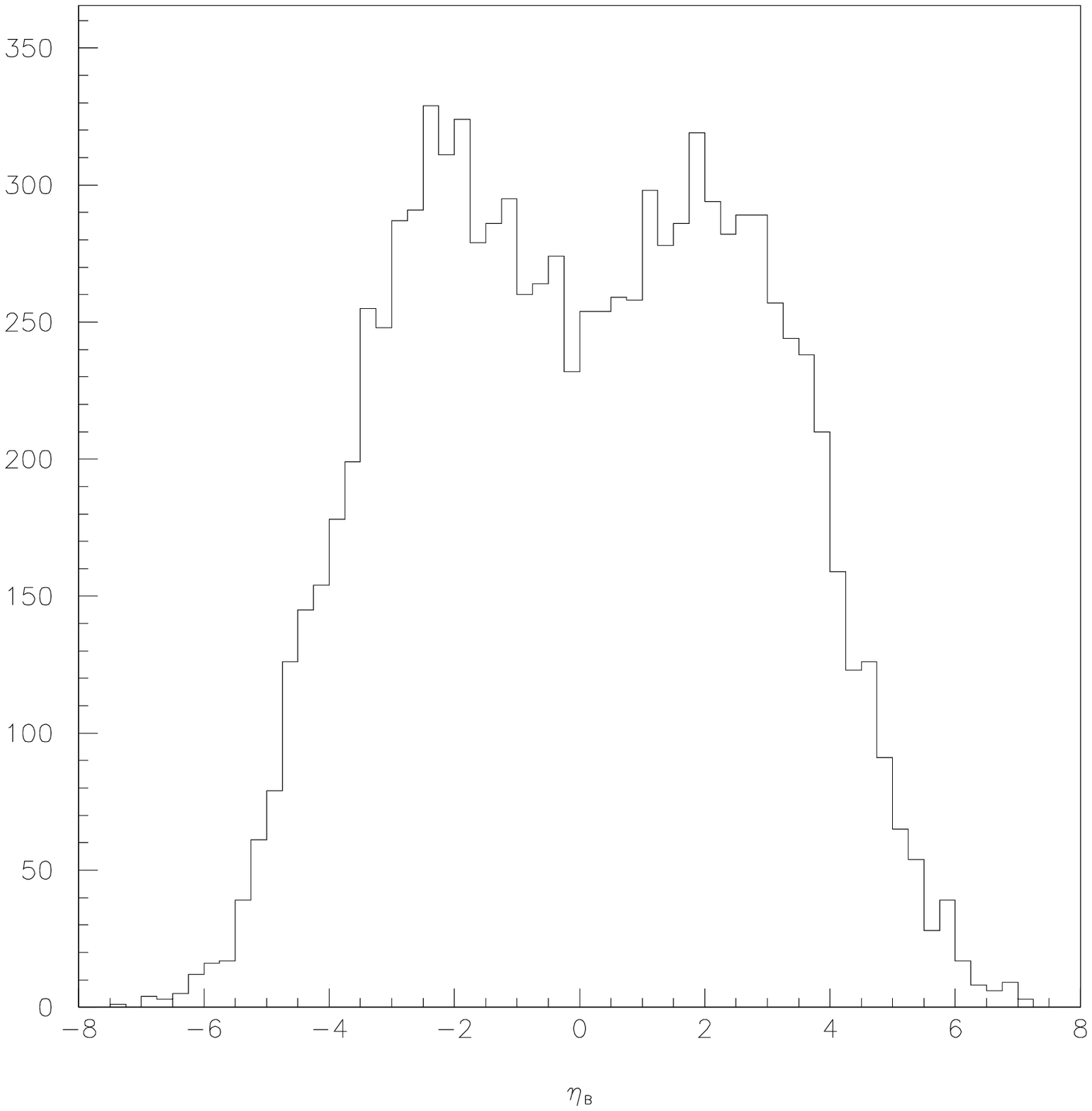,width=2.5in,bbllx=0bp,bblly=0bp,bburx=600bp,bbury=700bp,clip=}
\vspace{-9.15cm}\hspace*{2.4in}
\psfig{figure=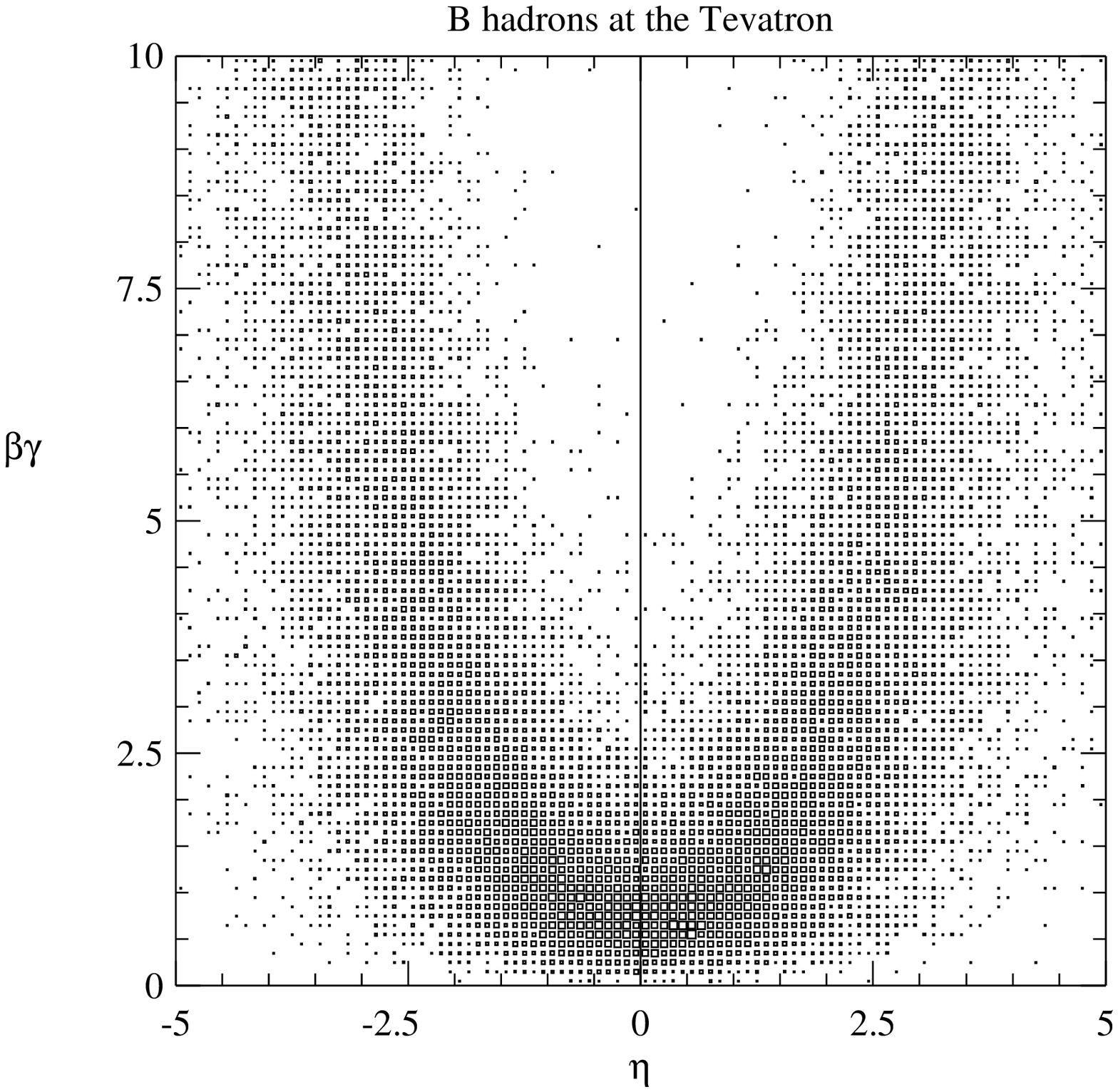,width=2.6in,bbllx=0bp,bblly=0bp,bburx=600bp,bbury=700bp,clip=}
\vspace{-.1cm}
\caption{\label{n_vs_eta}  The $B$ yield versus $\eta$ (left). 
$\beta\gamma$ of the  $B$  versus $\eta$ (right).}
\end{figure}

According to QCD based calculations of $b$ quark production, the $b$'s are
produced ``uniformly" in $\eta$ and have a truncated transverse momentum,
$p_t$, spectrum, characterized by a mean value approximately equal to the $B$
mass.\cite{artuso} The distribution in $\eta$ is shown in Fig.~\ref{n_vs_eta}.

\begin{figure}[htb]
\vspace{-1.4cm}
\centerline{\psfig{figure=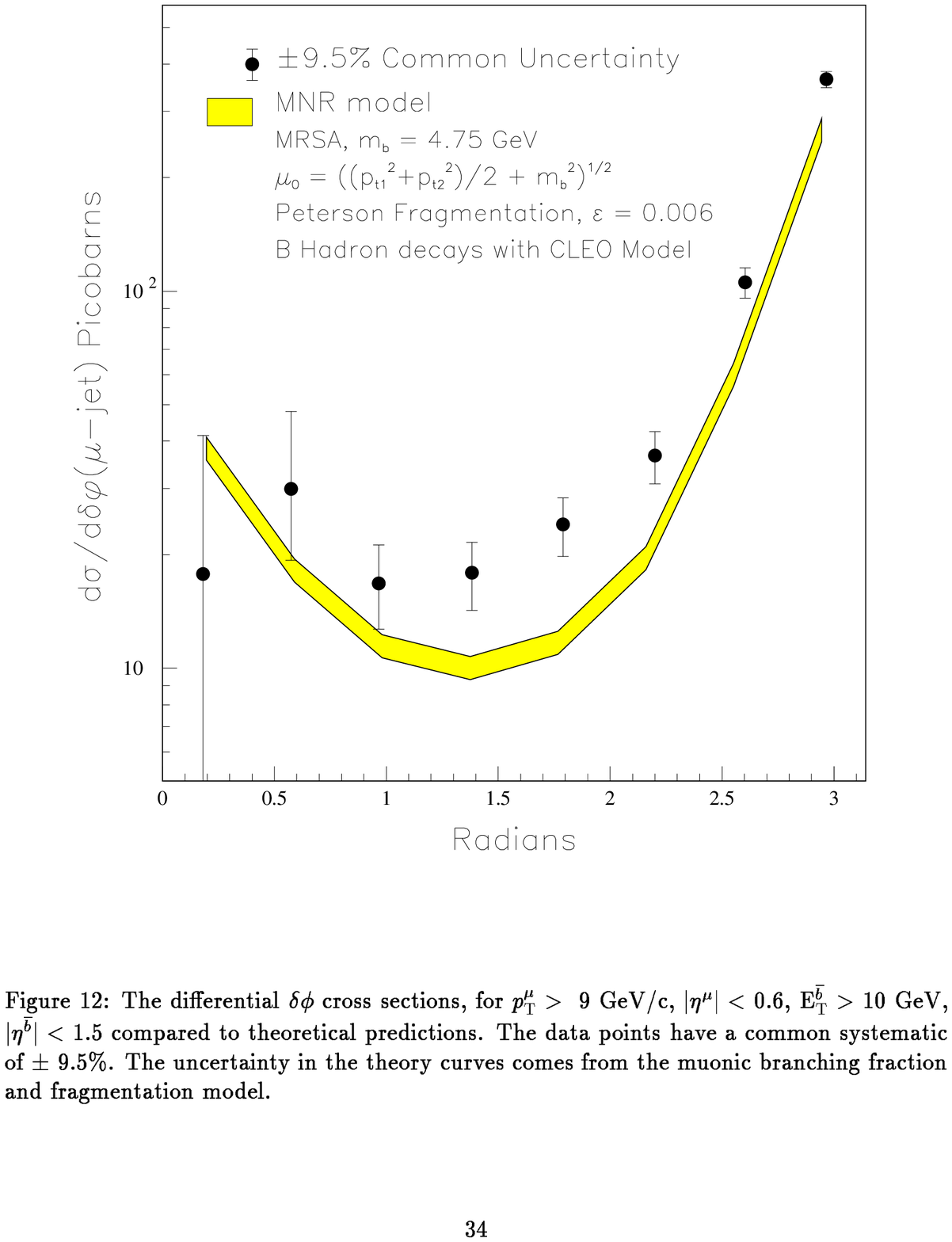,height=2.9in,bbllx=0bp,bblly=200bp,bburx=600bp,bbury=700bp,clip=}}
\vspace{-.5cm}
\caption{\label{cdf_dphi}The differential $\delta\phi$ cross-sections for
$p^{\mu}_T> 9 $ GeV/c, $\left|\eta^{\mu}\right|<$0.6, E$^{\bar{b}}_T>$10 GeV,
$\left|\eta^{\bar{b}}\right|<1.5$ compared with theoretical predictions. The
data points have a common systematic uncertainty of $\pm$9.5\%. The uncertainty
in the theory curve arises from the error on the muonic branching ratio and
the uncertainty in the fragmentation model.}
\end{figure}

There is a strong correlation between the $B$ momentum and $\eta$. Shown also in
Fig.~\ref{n_vs_eta} is the $\beta\gamma$ of the $B$ hadron versus $\eta$.
It can clearly be seen that near $\eta$ of zero, $\beta\gamma\approx 1$, while
at larger values of $|\eta |$, $\beta\gamma$ can easily reach values of ~6.
This is important because the observed decay length varies with $\beta\gamma$
and furthermore the absolute momenta of the decay products are larger allowing
for a supression of the multiple scattering error.

Since the detector design is somewhat dependent on the Monte Carlo generated
$b$ production distributions, it is important to check that the correlations
between the $b$ and the $\overline{b}$ are adequately reproduced. 
 In Fig.~\ref{cdf_dphi} I show the 
azimuthal opening angle distribution between a muon from a $b$ quark decay and the
$\bar{b}$ jet as measured by CDF~\cite{cdf_prod} and 
compare with the MNR predictions.\cite{MNR}
 
The model does a good job in
representing the shape which shows a strong back-to-back correlation. The
normalization is about a factor of two higher in the data than the theory,
which is generally
true of CDF $b$ cross-section measurements.\cite{cdf_bx} 
In hadron colliders all $B$ species are produced at the same time.

\begin{figure}[htb]
\vspace{-.4cm}
\centerline{\psfig{figure=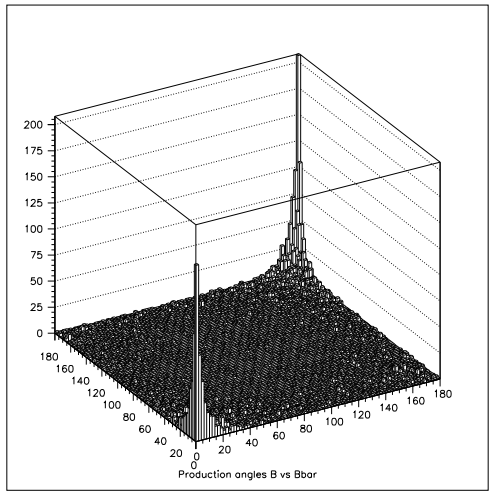,height=3.2in,bbllx=0bp,bblly=0bp,bburx=600bp,bbury=700bp,clip=}}
\vspace{-1.1cm}
\caption{\label{bbar}The production angle (in degrees) for the hadron
containing a $b$ quark plotted versus the production angle for a hadron
containing $\bar{b}$ quark.}
\end{figure}

The ``flat" $\eta$ distribution hides an important correlation of
$b\bar{b}$ production at hadronic colliders. In Fig.~\ref{bbar} the production
angles of the hadron containing the $b$ quark is plotted versus the production
angle of the hadron containing the $\bar{b}$ quark according to the Pythia
generator. There is a very strong
correlation in the forward (and backward) direction: when the $B$ is forward
the $\overline{B}$ is also forward. This correlation is not present in the
central region (near zero degrees). By instrumenting a relative small region of
angular phase space, a large number of $b\bar{b}$ pairs can be detected.
Furthermore the $B$'s populating the forward and backward regions have large
values of $\beta\gamma$.

Charm production is similar to $b$ production but much larger. Current
theoretical esitmates are that charm is 1-2\% of the total $p\bar{p}$
cross-section. Table ~\ref{tab:b_c} gives the relevant Tevatron parameters.
We expect to start serious data taking in Fermilab Run II with a luminosity of
about $5\times 10^{31}$cm$^{-2}$s$^{-1}$; our ultimate luminosity goal, to be
obtained in Run III is $2\times 10^{32}$cm$^{-2}$s$^{-1}$. 
\begin{table}[t]
\caption{The Tevatron as a $b$ and $c$ source for C0 in Run II.\label{tab:b_c}}
\vspace{0.4cm}
\begin{center}

\begin{tabular}{|l|c|}   \hline
Luminosity in Run II& $5\times 10^{31}$cm$^{-2}$s$^{-1}$\\
Luminosity (ultimate) & $2\times 10^{32}$cm$^{-2}$s$^{-1}$\\
$b$ cross-section & 100$\mu$b \\ 
\# of $b$'s per 10$^7$ sec  & $10^{11}$\\
$b$ fraction & 0.2\% \\
$c$ cross-section & $>500~\mu$b \\
Bunch spacing & 132 ns \\
Luminous region length & $\sigma_z$ = 30 cm\\
Luminous region length & $\sigma_x$ $\sigma_y$ = $\approx 50$ $\mu$m\\
Interactions/crossing & $<0.5>$
 \\ \hline
\end{tabular}
\end{center}
\end{table}

\section{The Experimental Technique: A Forward Two-arm Spectrometer}

\begin{figure}[htbp]
\vspace{.03cm}
{\psfig{figure=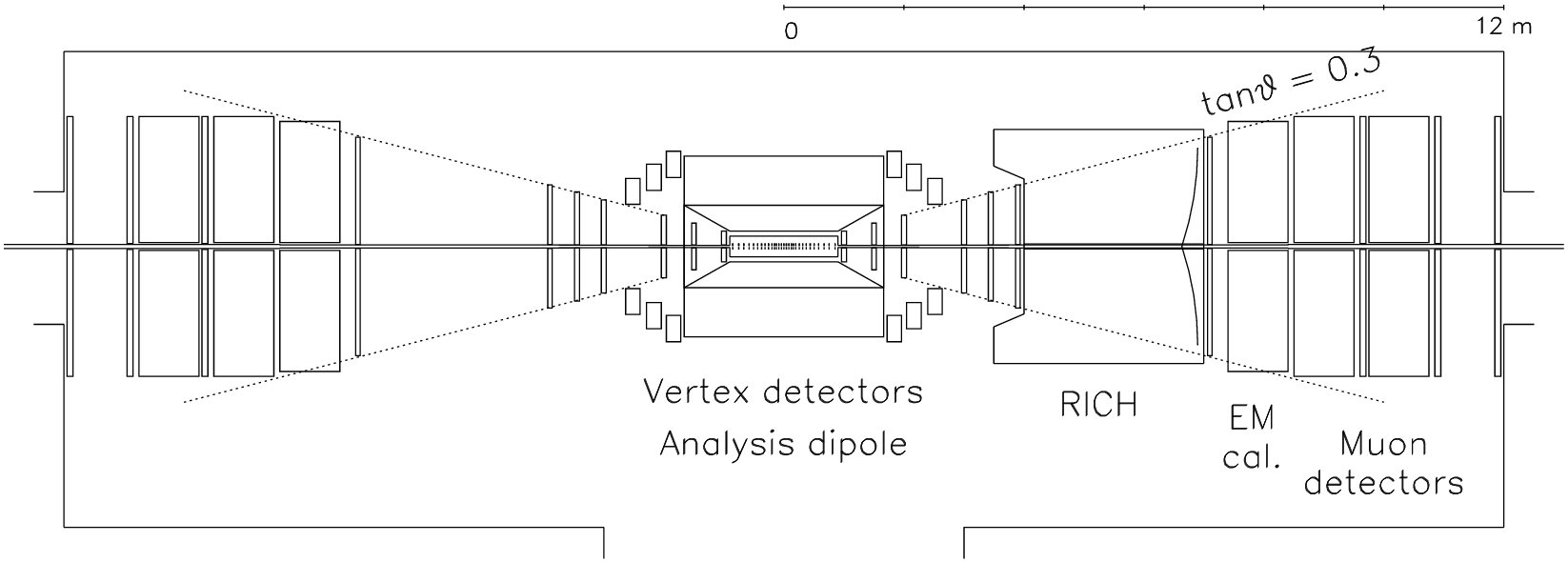,height=2.4in,bbllx=0bp,bblly=0bp,bburx=600bp,bbury=400bp,clip=}}
\vspace{-1.80cm}
\caption{\label{btev_det_doc}Sketch of the
BTeV spectrometer.}
\end{figure}

A sketch of the apparatus is shown in Fig.~\ref{btev_det_doc}. The plan view
shows the two-arm spectrometer fitting in the expanded C0 interaction region
at Fermilab. There is a construction money in the budget for excavating the
interaction region and installing a counting room. The magnet that
we will use, called SM3, exists at Fermilab. The other important parts of the
experiment include the vertex detector, the RICH detectors, the EM calorimeters
and the muon system.  

The solid angle subtended is approximately $\pm$300 mr in both plan and
elevation views. The vertex detector is a multiplane pixel device which sits
inside the beam pipe. The triggering concept is to pipleline the data and to
trigger on detached $b$ or $c$ verticies in the first trigger level. The vertex
detector is put in the magnetic field in order to insure that the tracks
considered for vertex based triggers do not have large mulitple scattering
because they are low momentum.

\section{Simulations}

We have developed several fast simulation packages to verify the basic BTeV
concepts and aid in the final design. The key program in our system is
MCFast.\cite{MCFast} Charged tracks are generated and traced through different
material volumes including detector resolution, multiple scattering and
efficiency. This allows us to measure acceptances and resolutions in a fast
reliable manner.

\begin{figure}[hbt]
\vspace{-.3in}
\psfig{bbllx=0pt,bblly=0pt,bburx=600pt,bbury=600pt,file=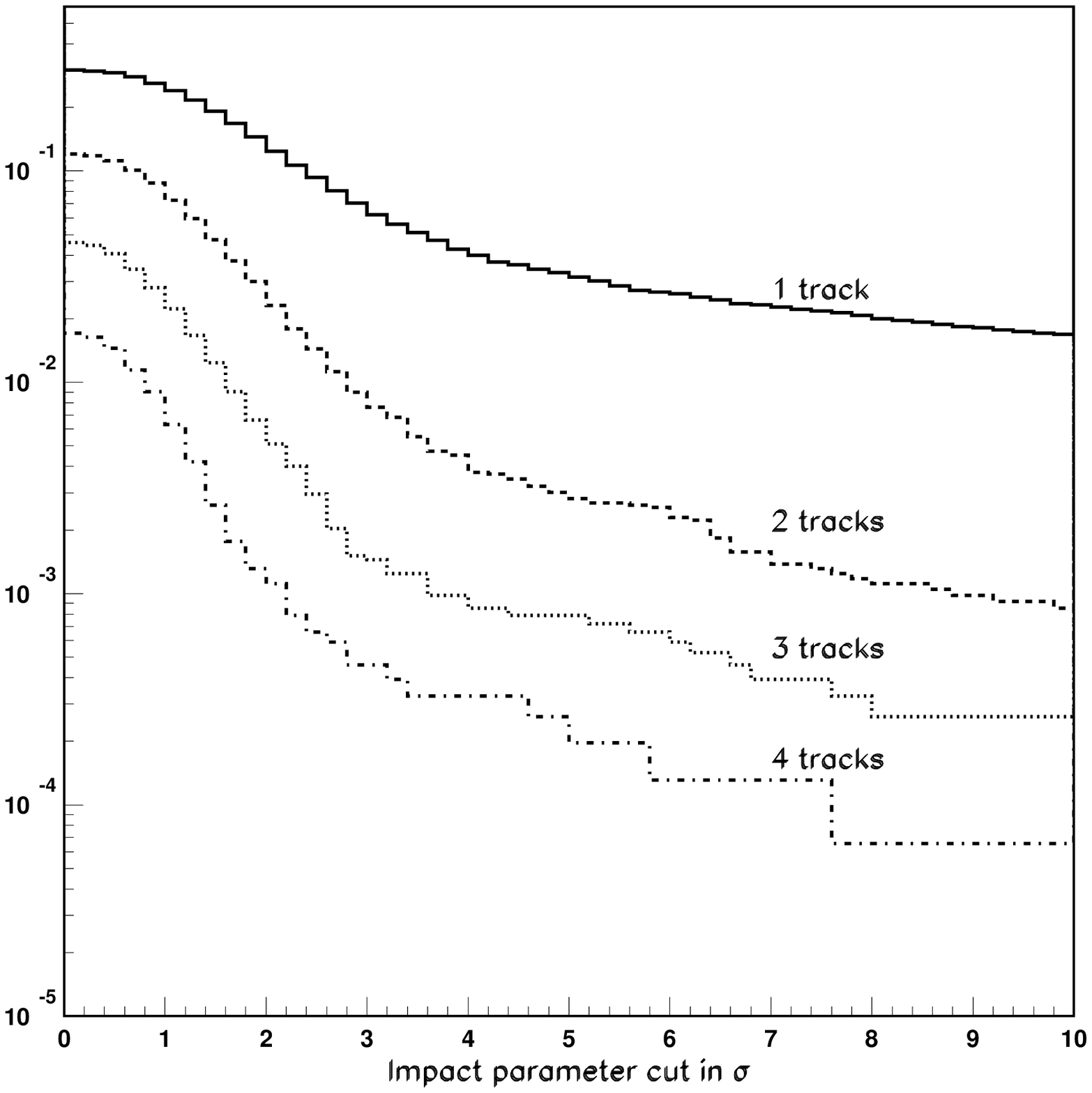,width=2.8in}
\vspace{-7.15cm}\hspace*{2.3in}
\psfig{bbllx=0pt,bblly=0pt,bburx=600pt,bbury=600pt,file=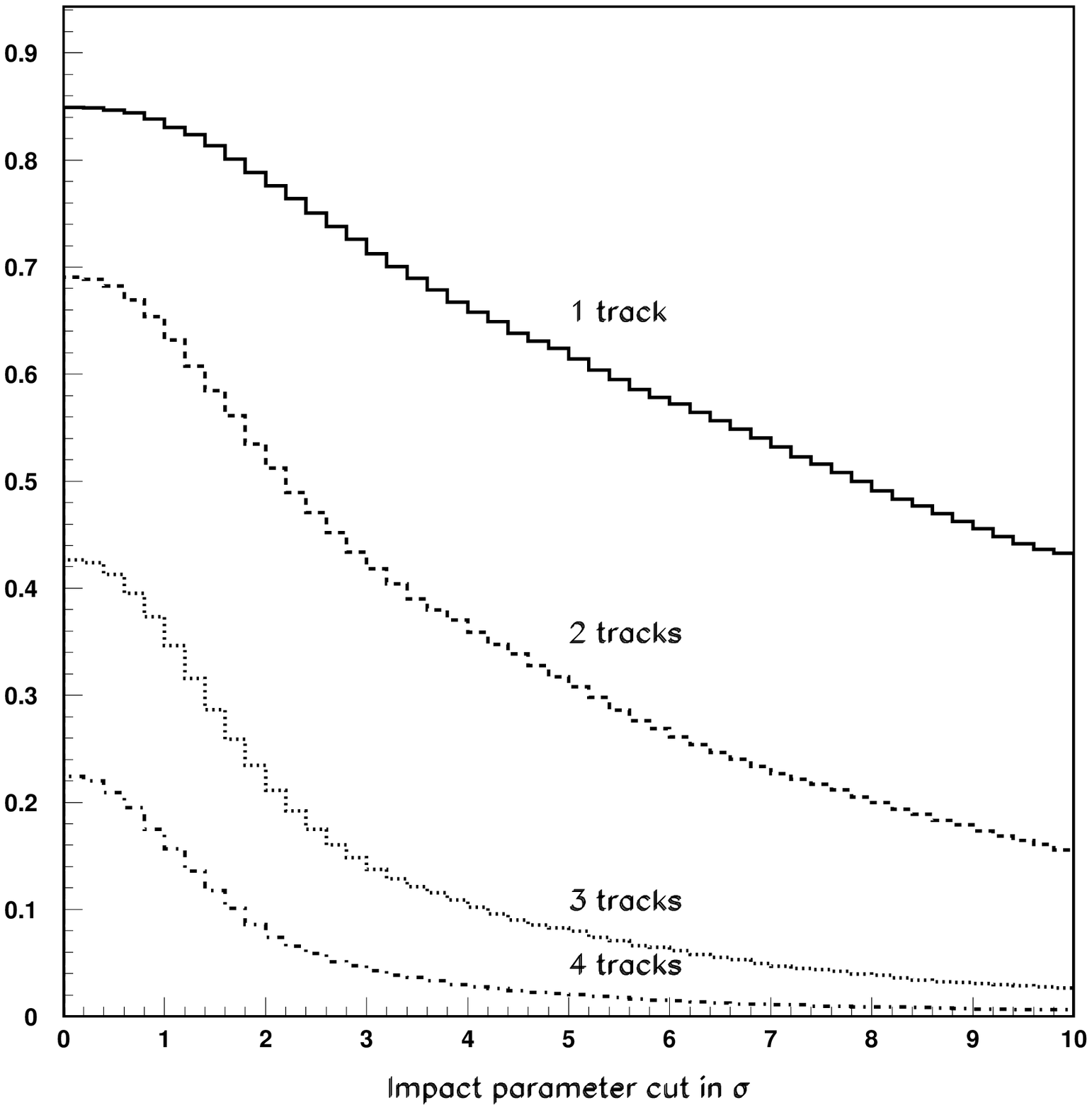,width=2.8in} 
\vspace{-.3cm}
\caption{\label{trig1} (left) Trigger efficiency for light quark events.
(right) Trigger efficiency for $B^o\to\pi^+\pi^-$ for pion tracks  in the spectrometer. The ordinate gives the choice of cut value
on the impact parameter in terms of number of standard deviations ($\sigma$) 
of the track from the primary vertex. The curves show the effect of requiring
 different numbers of tracks. }

\end{figure}

Our baseline trigger algorithm works by first determining the main event vertex
and then finding how many tracks miss this vertex by $n\sigma$, where $\sigma$
refers to the impact parameter divided by its error. Furthermore, a requirement
is then placed on the track momentum in the bend plane, $p_y$, as determined on
line. The preliminary results of simulating this algorithm are shown in
Fig.~\ref{trig1} for a cut $p_y~>~0.5$ GeV/c.\cite{Selove}
The choice
of the number of tracks and the impact parameter requirement must eventually 
be fixed, but what is shown here (left) is the efficiency for accepting light
quark events ($u$, $d$, and $s$) for various choices on the number of tracks
(curves) and the size of their required impact parameter divided by the 
error in impact parameter.
The efficiency for accepting  $B^o\to\pi^+\pi^-$ is shown in the right side.
Here the efficiency is given after requiring that both tracks are in the 
spectrometer and accepted for further analysis. For a ``typical" $n\sigma$ cut
of 3 and track requirement of 2, the $\pi^+\pi^-$ trigger efficiency is about 45\%,
while  The light
quark background  has an efficiency of about 0.8\%.
 Note, that we do not consider $c$ to be a background
in this experiment. For ``typical" charm reaction the same trigger gives about
a 1\% efficiency on charm.

\begin{figure}[htb]
\vspace{-1.8cm}
\centerline{\psfig{figure=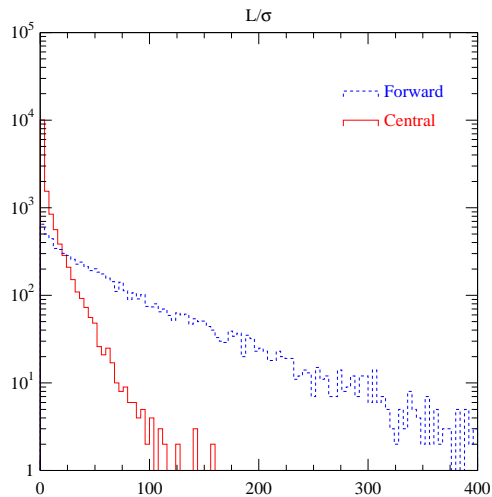,height=3.4in,bbllx=0bp,bblly=0bp,bburx=600bp,bbury=630bp,clip=}}
\vspace{-0.3cm}
\caption{\label{l_over_sig} Comparison of the $L/\sigma$ distributions
for the decay $B^o\to\pi^+\pi^-$ in central and forward detectors
produced at a hadron collider with a center of mass energy of 1.8~TeV.}
\vspace{1 cm}
\end{figure}

For the $B^o\to\pi^+\pi^-$ channel we have also compared the offline fully 
reconstructed decay length distributions in our forward geometry with that of
detector configured to work in the central region using MCFast. 
In Fig.~\ref{l_over_sig} I show the normalized 
decay length expressed in terms of $L/\sigma$ where $L$ is the decay length and 
$\sigma$ is the error on $L$ for the $B^o\to \pi^+\pi^-$ decay.\cite{procario}

 The forward 
detector clearly has a much more favorable $L/\sigma$ distribution, which is
due to the excellent proper time resolution. This will be crucial for studies
of $B_s$ mixing. For the decay mode $B_s^o\to \psi K^{*o}$ our proper time
resolution is 45 fs, which allows us to measure values of $x_s$ up to 50 in
several years of running.

We have also investigated the feasiblity of tagging kaons using a gas Ring
Imaging Cherenkov Counter (RICH) in a forward geometry and compared it with
what is possible in a central geometry using Time-of-Flight counters with
good, 100 ps, resolution. For the forward detector the momentum coverage
required is between 3 and 70 GeV/c. The lower momentum value is determined by
our desire to tag charged kaons for mixing and CP violation measurements, while
the upper limit comes from distingushing the final states $\pi^+\pi^-$,
$K^+\pi^-$ and $K^+K^-$. 
The momentum range is much lower in the central detector
but does have a long tail out to about 5 GeV/c.  Either C$_4$F$_{10}$ or
C$_5$F$_{12}$ have pion thresholds of about 2.5 GeV/c. The  kaon and proton
thresholds for the first gas are 9 and 17 GeV/c, respectively. 

\begin{figure}[htb]
\vspace{-1.3cm}
{\psfig{figure=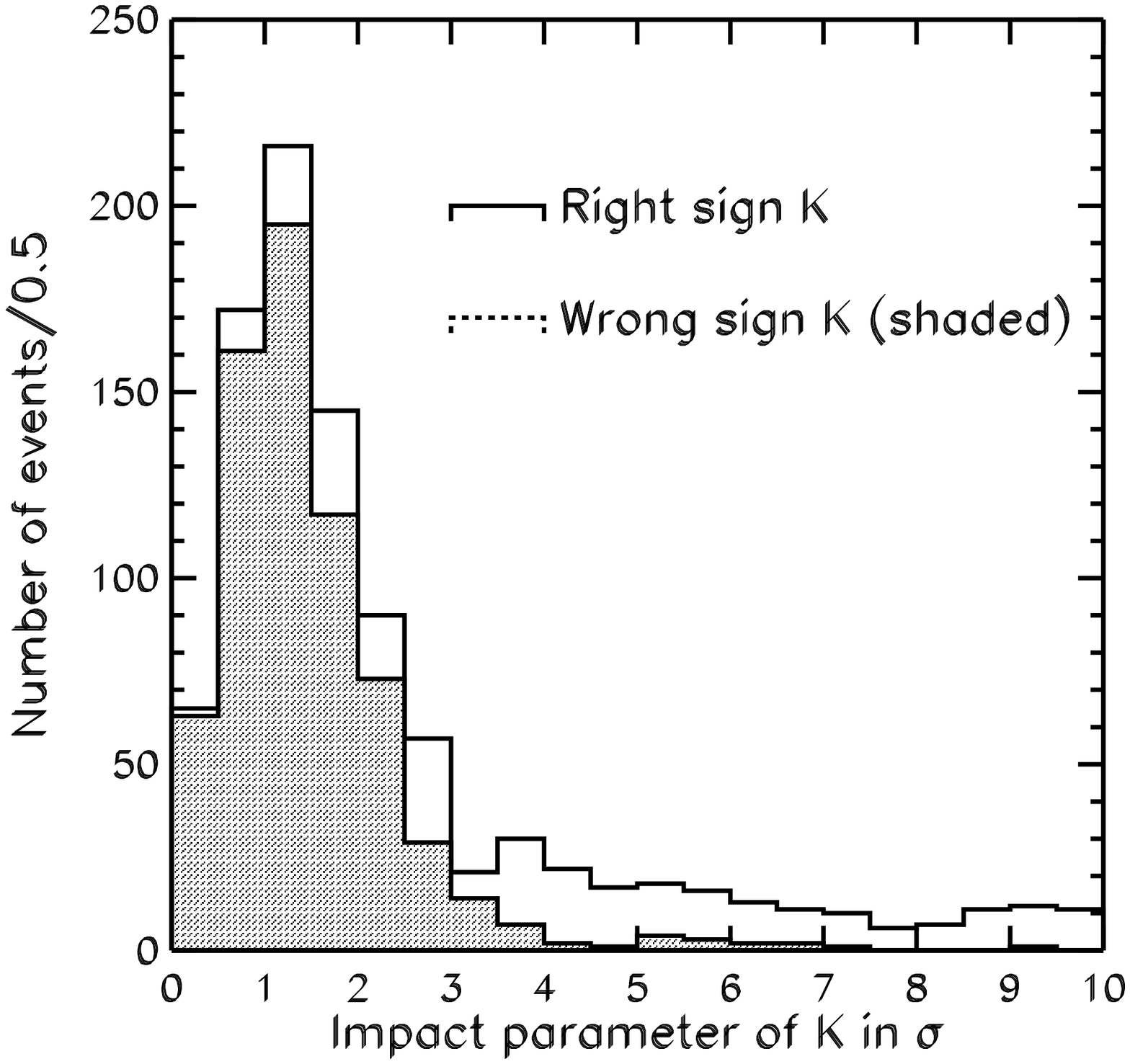,height=2.8in,bbllx=0bp,bblly=0bp,bburx=600bp,bbury=613bp,clip=}}
\vspace{-7.2cm}\hspace*{2.4in}
\psfig{figure=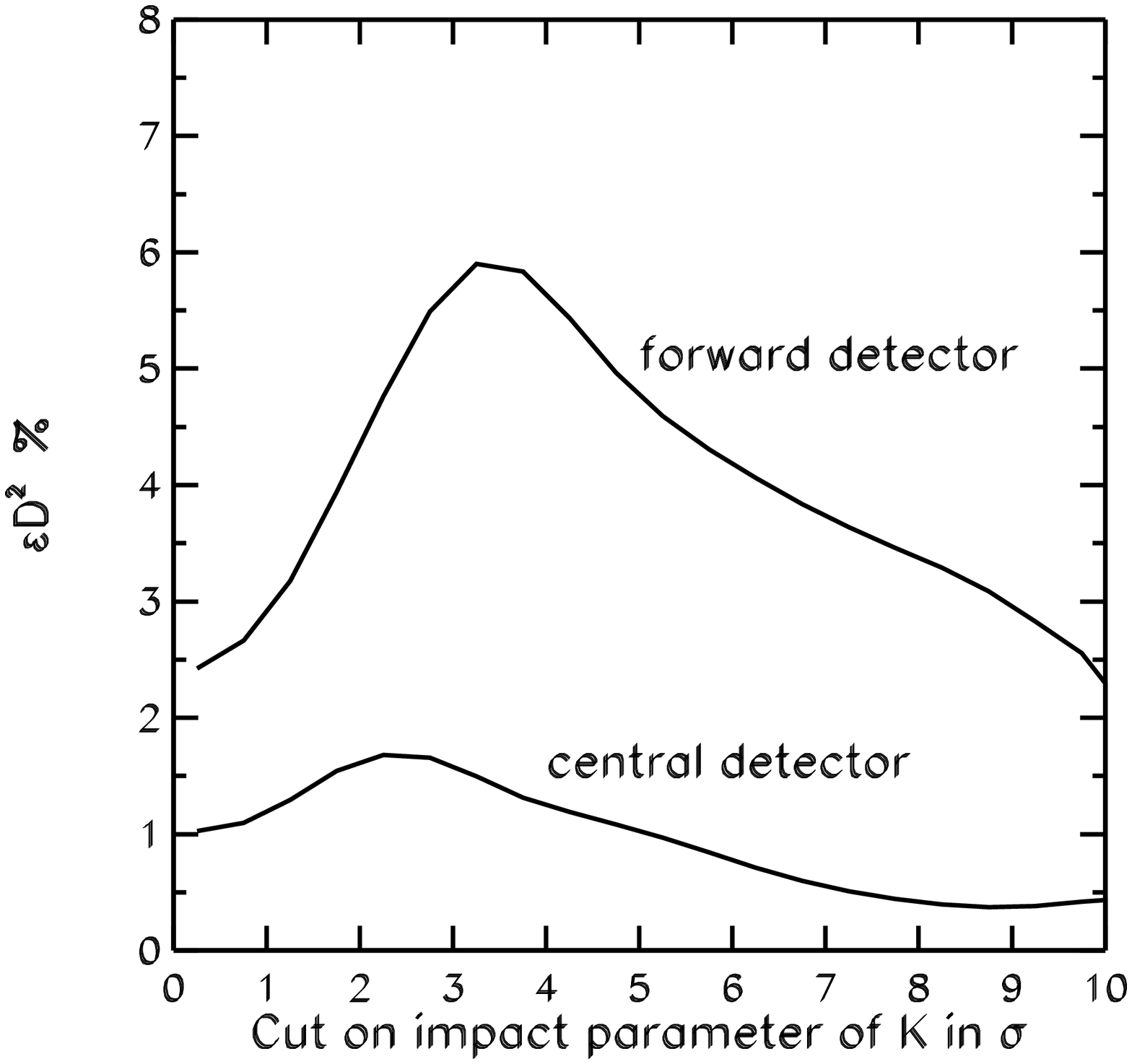,height=2.8in,bbllx=0bp,bblly=0bp,bburx=600bp,bbury=630bp,clip=}
\caption{\label{ipk}(left) $L/\sigma$ distributions in BTeV for $K^{\pm}$ impact
parameters for right sign (unshaded) and wrong sign (shaded) tags. (right) Overall $\epsilon$D$^2$ values from kaon tagging for a forward detector
containing a flourine based RICH versus a central detector with 100 ns time of
flight resolution as a function of kaon impact parameter in units of
$L/\sigma$.
(Protons have been ignored in both cases.)}
\end{figure}

The BTeV RICH was simulated using the current C0 geometry with MCFast.
Fig.~\ref{ipk} shows the number of identified kaons plotted versus their
impact parameter divided by the error in the impact parameter for both right
sign and wrong sign kaons. A right sign kaon is a kaon which properly tags the
flavor of the other $B$ at production. We expect some wrong sign kaons from
mixing and charm decays. Many others just come from the primary. A cut 
on the impact parameter standard deviation plot at $3.5\sigma$ gives an overall
$\epsilon D^2$ of 6\%. Here $\epsilon$ is the efficiency and $D$ is the
dilution which can be expressed as number of right sign minus the number of
wrong sign divided by the sum. The 6\% may be an over-estimate because protons
can contribute to the sample. Putting these in lowers $\epsilon D^2$ to 5.1\%.
These numbers are for a perfect RICH system. Putting in a fake rate of several
percent, however, does not significantly change this number.

The simulation of the central detector gives much poorer numbers. In
Fig.~\ref{ipk}
$\epsilon D^2$ for both the forward and central detectors are shown as a
function of the kaon impact parameter (protons have been ignored). It is
difficult to get $\epsilon D^2$ of more than 1.5\% in the central detector. This
analysis showed the importance of detecting protons so we are now in the
process of seeing if we can provide low momentum K/p separation using another
radiator.

\section{Conclusions}
Progress has been made toward starting a program to measure mixing, CP
violation and rare decays in both the $b$ and $c$ systems at the Fermilab
collider. The pit at the C0 collision hall is being enlarged and can accomodate
a two-arm forward spectrometer as described above. The BTEV collaboration hopes
to start taking data, with at least a portion of the apparatus during collider
Run II at about the same time the Tevatron changes to 132 ns bunch spacing,
although test runs are expected well before. The full spectrometer is hoped for
in collider Run III. The detector design encorporates detached vertex
triggering at the first level with excellent tracking and charged particle
identification using a Ring Imaging Cherenkov system. It also has separate
elements for electron and muon detection.\cite{web} 

The enormous potential physics power of the experiment is reflected by how well
we can expect to measure the CP violating asymmetry in the decay $B^o\to
\pi^+\pi^-$. Table ~\ref{tab:pipi} gives the relevant parameters. In one year
of ``low" luminosity running
the asymmetry can be measured to an accuracy of $\pm$0.05.

\begin{table}[t]
\caption{The Tevatron as a $b$ and $c$ source for C0 in Run II.\label{tab:pipi}}
\vspace{0.4cm}
\begin{center}

\begin{tabular}{|l|c|}   \hline
Luminosity & $5\times 10^{31}$cm$^{-2}$s$^{-1}$\\
$b$ cross-section & 100$\mu$b \\ 
Number of $B_d^o$'s & $3.5\times 10^{10}$ \\
${\cal B}(B_d^o\to\pi^+\pi^-)$ & 0.75$\times 10^{-5}$\\
Reconstruction efficiency & 0.09\\
Trigger efficiency & 0.72\\
Number reconstructed $\pi^+\pi^-$ & 1.7$\times 10^4$\\
Tagging Efficiency $\epsilon D^2$& 0.10\\
Signal/Background & 0.40\\
error in asymmetry & 0.05
 \\ \hline
\end{tabular}
\end{center}
\end{table}

\section*{Acknowledgments}
I would like to thank my BTeV colleagues including Joel Butler, Chuck Brown
and Patty McBride, Tomasz Skwarnicki and Kevin Sterner for their help in 
getting this material together. 

\end{document}